\pdfoutput=1

\documentclass[8.5pt,twoside,twocolumn]{article}
\usepackage[super,sort&compress,comma]{natbib} 
\usepackage{mhchem}
\usepackage{times,mathptmx}
\usepackage{sectsty}
\usepackage{balance} 

\usepackage{graphicx} %eps figures can be used instead
\usepackage{lastpage}
\usepackage[format=plain,justification=raggedright,singlelinecheck=false,font=small,labelfont=bf,labelsep=space]{caption} 
\usepackage{fancyhdr}
\usepackage{comment}
\usepackage{color}

\newcommand{\We}{\mbox{We}}
\newcommand{\lb}{\left(}
\newcommand{\rb}{\right)}
\newcommand{\lsb}{\left[}
\newcommand{\rsb}{\right]}

\title{A solvable model of axisymmetric and non-axisymmetric droplet bouncing}
\author{Matthew Andrew\footnote{\textit{~The Rudolf Peierls Centre for Theoretical Physics, 1 Keble Road, Oxford, OX1 3NP, UK. E-mail: julia.yeomans@physics.ox.ac.uk}}, Julia M. Yeomans$^*$ and Dmitri O. Pushkin\footnote{\textit{~Department of Mathematics, University of York, York, YO10 5DD, UK. E-mail: mitya.pushkin@york.ac.uk}}}

\begin{document}

\maketitle

\abstract{We introduce a solvable Lagrangian model for droplet bouncing. The model predicts that, for an axisymmetric drop, the contact time decreases to a constant value with increasing Weber number, in qualitative agreement with experiments, because the system is well approximated as a simple harmonic oscillator. We introduce asymmetries in the velocity, initial droplet shape, and contact line drag acting on the droplet and show that asymmetry can often lead to a reduced contact time and lift-off in an elongated shape. The model allows us to explain the mechanisms behind non-axisymmetric bouncing in terms of surface tension forces. 
Once the drop has an elliptical footprint the surface tension force acting on the longer sides is greater. Therefore the shorter axis retracts faster and, due to the incompressibility constraints, pumps fluid along the more extended droplet axis. This leads to a positive feedback, allowing the drop to jump in an elongated configuration, and more quickly.}

\section{Introduction}

%\jucomment{Mitya mentioned physical relevance of chaotic mode? hard to see for a free drop but maybe would we see it on a surface? - is this what the lotus drops were doing?}

The interaction of water droplets with solid surfaces is of importance to a wide range of applications including ink-jet printing\cite{singh2010inkjet}, spray cooling\cite{kim2007spray}, ice accumulation\cite{jung2011superhydrophobic, hejazi2013superhydrophobicity} and soil erosion by rainfall\cite{kinnell2005raindrop}. The impact process can be complex: Depending on their size, impact velocity, and the nature of the surface, drops can be deposited on the surface, break-up and splash, or bounce\cite{josserand2016drop, yarin2006drop, marengo2011drop}.

When a drop lands on a solid surface inertial forces mediated by the contact with the surface cause the drop to spread out laterally. As it does so, its kinetic energy is transformed into surface energy. The fluid comes to rest and the stored surface energy causes the drop to retract towards its original spherical shape. If it does so with enough energy it will rebound from the surface. The timescale associated with the bouncing follows from a scaling argument balancing inertia and surface tension as $\tau\sim(\rho R^3 / \sigma)^{\frac{1}{2}}$ where $\rho$ is the density, $R$ is the radius and $\sigma$ is the surface tension of the drop.

Superhydrophobic surfaces are characterised by high contact angles and low contact angle hysteresis \cite{shirtcliffe2010introduction, quere2008non, chen2011comparative, bartolo2006bouncing, wang2007impact,reyssat2010dynamical}. Richard {\it et al.}$^{\cite{richard2002surface}}$ performed experiments showing that the contact time of  a bouncing drop on a superhydrophobic surface is $2.6 \tau$ for high enough impact speeds, and that viscosity is not important in some regimes of droplet bouncing. Almost elastic collisions can also be achieved on a Leidenfrost surface or if a trapped air layer is preserved below the drop\cite{quere2013leidenfrost, biance2011drop}. 

One theoretical approach to describing drop bouncing is in terms of the normal modes of vibration. In a classic paper Rayleigh\cite{rayleigh1879capillary} calculated the period of small oscillations in the shape of a drop about the spherical equilibrium as $2.2 \tau$. Courty {\it et al.}\cite{courty2006oscillating} extended this work to drops at a surface. They found that introducing a surface increased the oscillation period compared to free oscillations and, assuming that the contact time can be viewed as half an oscillation period of the lowest frequency harmonic, predicted a contact time of $2.3 \tau$. 

More recently several authors have described droplet bouncing on surfaces that lack isotropic symmetry. Examples include micro-scale ridges on a flat surface\cite{bird2013reducing}, superhydrophobic stripes\cite{song2015selectively}, cylindrical substrates\cite{liu2015symmetry}, and wires laid upon surfaces\cite{gauthier2015water}.
These experiments and simulations showed that inducing non-axisymmetric bouncing modes reduces the contact time of a drop on a surface below that found for axisymmetric collisions.

In Sec.~\ref{sec2} we introduce a simple model of drop bouncing. Our model has the advantage over the Rayleigh approach in that it does not assume small deformations of the drop and so can go beyond linearity. In Sec.~\ref{sec3} we present our results. We consider the axisymmetric and non-axisymmetric motion of a free drop, showing that the drop oscillates chaotically in the non-axisymmetric case. We then calculate the contact time of an axisymmetric bouncing drop, which decreases to a constant value with increasing Weber number, in qualitative agreement with experiments. Next asymmetries in velocity, initial droplet shape, or drag are introduced. We show that asymmetry often leads to a reduced contact time and lift-off in an elongated shape, and we use analytical arguments and numerical solutions of the governing equations to explain why this is the case.

%The motivation for doing so is to investigate several different ways in which a lateral asymmetry can affect the bouncing of drops. Our model has the advantage over the Rayleigh approach in that it does not assume only small deformations of the drop so can go beyond linearity. This allows a comparison to be made to experimental results and a qualitative match to be found. The model can then be used to study bouncing with introduced asymmetry finding that this can lead to a reduced contact time and lift-off in an elongated shape under the right conditions.

%\ju{discuss whether we should use axisymmetric, symmetric \ asymmetric, non-axisymmetric}
\section{The droplet model}
\label{sec2}

\subsection{The free drop}

We introduce a simple model which reproduces many of the features of droplet bouncing. Our first assumption is to neglect viscous dissipation in the fluid and assume zero friction with the surface. This means that the system is conservative and hence can be described by a Lagrangian.
 Secondly, we assume that the drop always takes an ellipsoidal shape which can be characterised by its three axes, of lengths $a$, $b$, $c$ along the $x$-, $y$- and $z$-directions respectively.  
Hence its volume, which is a conserved quantity, is $V_0 ={4 \pi abc }/{3}$. A convenient choice of fluid velocity, corresponding to an irrotational flow of incompressible fluid within the drop, allows the problem to be recast in terms of the evolution of the lengths of the axes of the ellipsoid:
\begin{equation}
\mathbf{u}=\left( x\frac{\dot{a}}{a},\,y\frac{\dot{b}}{b},\,z\frac{\dot{c}}{c} \right).
\end{equation}

There are two contributions to the Lagrangian describing the drop, the kinetic energy and the potential energy. The kinetic energy, $T$, follows by integrating over the volume of the ellipsoid. The choice of origin for this integration determines the centre of mass motion of the drop; here we take the origin to be the centre of the ellipsoid, corresponding to no centre of mass motion, and giving
\begin{equation}
T=\int \frac{\rho}{2} \lb u_x^2+u_y^2+u_z^2 \rb dV=\frac{\rho V_0}{2} I \lb \dot{a}^2+\dot{b}^2+\dot{c}^2 \rb
\label{eq:freeT}
\end{equation}
where $\rho$ is the density of the fluid of the drop and $I=1/5$ is the numerical factor associated with the moment of inertia of the ellipsoid.

We consider drops smaller than the capillary length and neglect gravity. Therefore the only contribution to the potential energy of the drop arises from the surface tension, $\sigma$, and is proportional to the surface area of the ellipsoid. As the surface area is in general given by elliptical integrals an approximation is useful:
\begin{equation}
U=4 \pi\sigma \left( \frac{(ab)^{\alpha}+(bc)^{\alpha}+(ac)^{\alpha}}{3}\right)^{1/\alpha} \quad \alpha = 1.6. 
\end{equation}
For this value of $\alpha$ the formula gives a relative error of less that 1.42$\%$  for all ellipsoid shapes. 

The Lagrangian of the system is 
\begin{equation}
\mathcal{L}=T-U +p \lb \frac{4 \pi abc}{3}-V_0 \rb,
\label{Lagrangian}
\end{equation}
where the final term is a Lagrange multiplier added to enforce the incompressibility constraint; the physical meaning of $p$ is pressure. In the system of units with length measured in units of the drop radius $R$ and time measured in units of $\tau_{s}=(\rho R^3/\sigma)^{1/2}$ the Lagrangian of the system can be written as 
\begin{equation}
\mathcal{L}=T(\dot{\mathbf{a}})-U(\mathbf{a}) + \tilde{p} \lb abc - 1 \rb,
\label{e:L}
\end{equation}
where $\mathbf{a}=(a,b,c)$, $\tilde{p}=pR/\sigma$ is the dimensionless pressure,
\begin{equation}
T(\dot{\mathbf{a}}) =T(\dot{a},\dot{b}, \dot{c}) = \frac{1}{2}\lb I{\dot{a}}^2 +  I {\dot{b}}^2 + I {\dot{c}}^2 \rb
\label{e:T}
\end{equation}
is a quadratic form of $\dot{\mathbf{a}}$ and 
\begin{equation}
U(\mathbf{a}) \approx 3^{1-1/\alpha} ( (ab)^\alpha +  (bc)^\alpha + (ca)^\alpha )^{1/\alpha},
\label{e:U}
\end{equation}
with $\alpha = 1.6$ is a homogeneous function of degree $2$. 

The corresponding Euler-Lagrange equations read
\begin{eqnarray}
I \ddot{a} &=& - \partial_a U + \tilde{p}/a, \label{e:Ia}\\
I \ddot{b} &=& - \partial_b U + \tilde{p}/b, \label{e:Ib}\\
I \ddot{c} &=& - \partial_c U + \tilde{p}/c, \label{e:Ic}\\
abc&=&1.\label{e:V0}
\end{eqnarray}
This dynamical system describes how fluid inertia, surface tension and pressure forces determine the dynamics of the bouncing drop.   

%  In the current system of units $E=\We/2+3$.

An elegant and useful expression for pressure $\tilde{p}$ can be obtained by multiplying Eq.~(\ref{e:Ia}) by $a$, Eq.~(\ref{e:Ib}) by $b$ and Eq.~(\ref{e:Ic}) by $c$, summing them, and transforming the result using the Euler identity for $U$
\[
\mathbf{a} \cdot \frac{\partial U}{\partial \mathbf{a}} = 2 U
\]
and the identities 
\[
 a \ddot{a}=\frac{1}{2} \frac{d^2a^2}{dt^2} - \dot{a}^2, \quad
 b \ddot{b}=\frac{1}{2} \frac{d^2b^2}{dt^2} - \dot{b}^2, \quad
 c \ddot{c}=\frac{1}{2} \frac{d^2c^2}{dt^2} - \dot{c}^2.
\]
Then,
\begin{equation}
\tilde{p}= \frac{1}{3} \lb  \frac{d^2T(\mathbf{a})}{dt^2} -2T(\dot{\mathbf{a}}) + 2 U (\mathbf{a}) \rb.
\label{e:p}
\end{equation}
In particular, when the drop is maintained in equilibrium the time derivatives vanish and an equivalent of the Young-Laplace law for capillary pressure is recovered:
\[
\tilde{p}_{\text{capillary}}= \frac{2}{3} U(\mathbf{a}).
\] 
Finally, we should notice that since the system conserves energy,
\begin{equation}
T(\dot{\mathbf{a}}) + U (\mathbf{a})=E,
\label{e:E}
\end{equation}
where $E$ is constant on any trajectory.

%can be rearranged to give
%\begin{eqnarray}
% p&=&\frac{\rho}{15}(a\ddot{a}+b\ddot{b}+c\ddot{c})+\frac{2 U}{3 V_0}, \label{eq:langmod1}\\
 %\ddot{a}a-\ddot{c}c&=&\frac{5}{\rho V_0}(c\partial_c U-a\partial_a U ), \label{eq:langmod2}\\
  % \ddot{b}b-\ddot{c}c&=&\frac{5}{\rho V_0}(c\partial_c U-b\partial_b U ). \label{eq:langmod3}
 %\end{eqnarray}
%Eq.~(\ref{eq:langmod1}) provides a value for the Lagrange multiplier $p$, which corresponds to the pressure. The first term on the rhs is a dynamical pressure and the second the Young-Laplace pressure. Eqs.~(\ref{eq:langmod1})--(\ref{eq:langmod3}) can be solved numerically as a closed system to model the oscillations of a free drop. 
%\end{comment}

\subsection{The bouncing drop}

The formalism can be extended to describe a drop hitting a flat surface at $z=-c$ by choosing a velocity field
\begin{equation}
\mathbf{u}=\left( x\frac{\dot{a}}{a},\;y\frac{\dot{b}}{b}\;,(z+c)\frac{\dot{c}}{c} \right).
\end{equation}
This model represents slip boundary conditions, since for a drop that initially lies above the surface the vertical component of velocity reaches zero at $z=-c$. Then, the free drop kinetic energy, Eq.~(\ref{e:T}), is replaced by   
\begin{equation}
T(\dot{\mathbf{a}}) =T(\dot{a},\dot{b}, \dot{c}) = \frac{1}{2} \lb I{\dot{a}}^2 +  I {\dot{b}}^2 +(I+1) {\dot{c}}^2 \rb
\label{eq:surfT}
\end{equation}
where the additional term in the kinetic energy is the energy associated with the motion of the centre of mass of the drop. Eq.~(\ref{e:Ic}) is replaced by
\begin{equation}
(I+1) \ddot{c} = - \partial_c U + \tilde{p}/c. \label{e:Ic2}\\
\end{equation}

For small amplitudes the drop oscillates without lifting off the surface.  At higher amplitudes oscillations do not occur, but instead the drop leaves the surface after a finite contact time. It is assumed that for the time prior to the drop lifting off the surface $\ddot{c}>0$ as the surface exerts a positive force on the drop. Therefore we identify the time of lift-off by the conditions  
\begin{equation}
\ddot{c}=0, \quad \dot{c}>0.
\label{cond:lift}
\end{equation} 

Contact line drag can be modelled by adding forcing terms of the form $F_a=-k_a \dot{a} b $ and $F_b=-k_b \dot{b} a $ to Eq.~(\ref{e:Ia}) and Eq.~(\ref{e:Ib}) respectively. This form is chosen so that the drag force is proportional to both the velocity in a given direction and to the length of the drop interface perpendicular to that direction. %\jucomment{is it OK to put friction, which is not a conservative force, into this Lagrangian formulation?}

\subsection{Initial Conditions}

The relevant variables in the model are $(a, b,c)$, describing the drop shape, and $(\dot{a}, \dot{b}, \dot{c})$, describing its velocity. 
The drop is initially chosen to be a sphere by setting $a_0=b_0=c_0=1$. To define the initial velocities we choose an initial kinetic energy $T_0$ corresponding to a Weber number 
\begin{equation}
\We=\frac{2 T_0  a_0}{V_0 \sigma}  
\end{equation}
where $V_0$ is the constant volume of the drop, and a value for $\gamma={\dot{a}_0}/{\dot{b}_0}$, the initial degree of lateral asymmetry in the velocities of the drop. The third initial velocity follows automatically from the constraint that the drop is incompressible. %\ju{OMIT at t=0 -- the drop is always incompressible}

If a surface is present the drop is assumed to be just touching the surface at $t=0$ and Eq.~(\ref{eq:surfT}) is used for the kinetic energy contribution to the Lagrangian.
Note that the constraint on incompressibility leads to an initial velocity that already has components in the transverse directions. Physically this models times after the initial crush phase of impact, which is a short but highly compressible regime, when incompressibility again becomes a good approximation\cite{eggers2010drop}.

In the system of units with length measured in units of $R$ and time measured in units of $\tau_{s}$ the total energy of the drop
\begin{equation}
E=T_0+U(1,1,1)=\We/2+3.
\end{equation}

\begin{figure*}
  \centering
  \includegraphics[trim = 0 0 0 0, clip, width=0.8\linewidth]{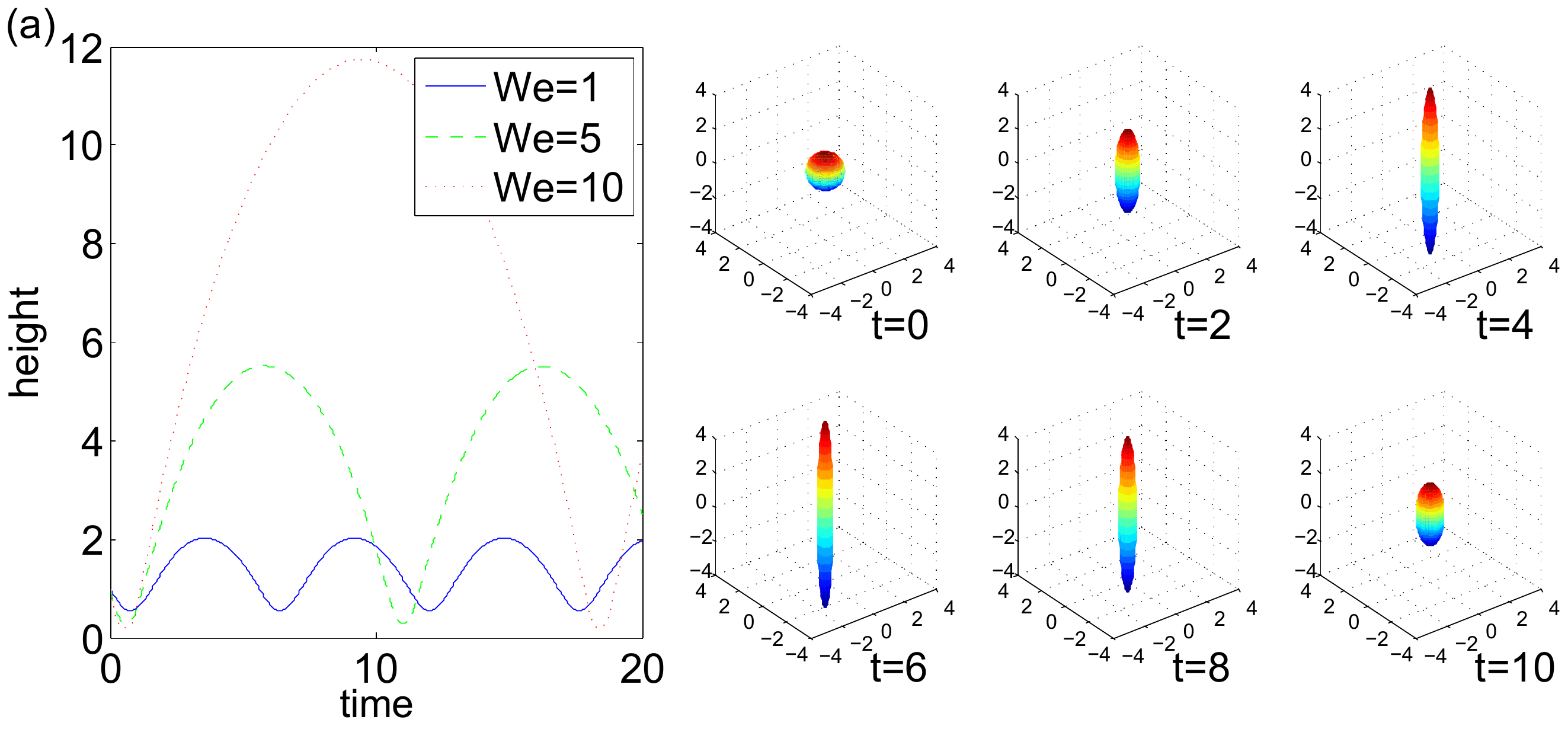}
   \centering
\includegraphics[trim = 0 0 0 0, clip, width=0.8\linewidth]{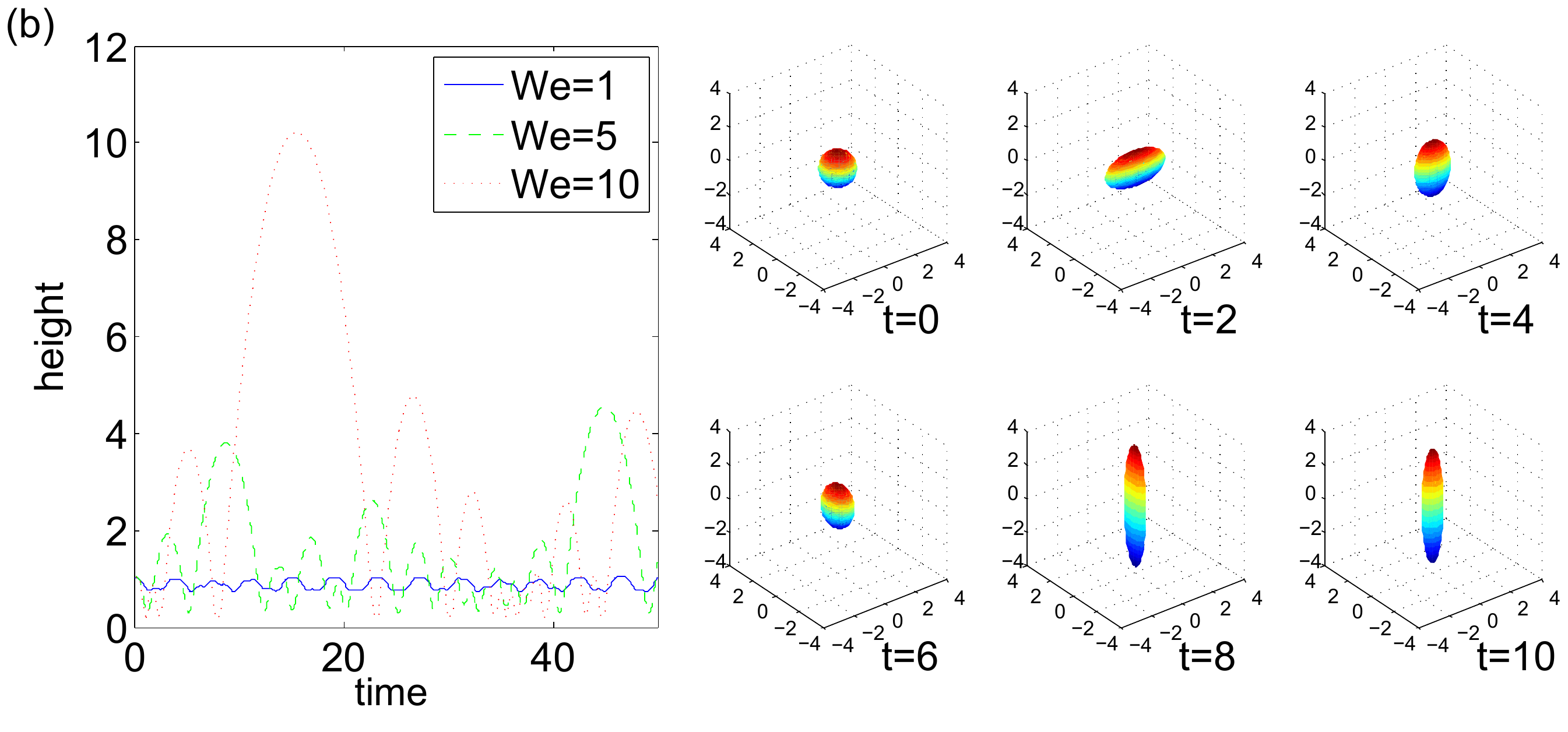}
 \centering
\includegraphics[trim = 0 20 0 0, clip, width=0.5\linewidth]{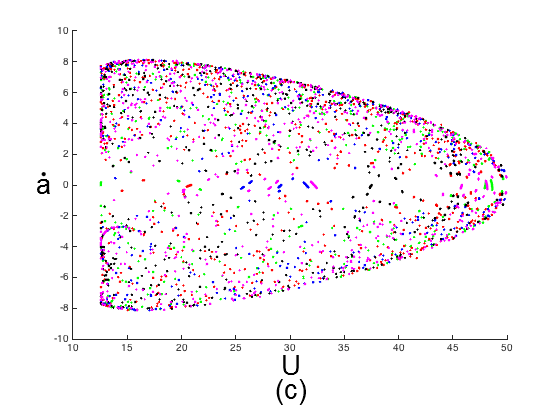}
  \caption{\label{fig:freeosc}(a) Time evolution of the vertical axis $c$ of a free drop oscillating (a) axisymmetrically (b) non-axisymmetrically  at $\We=1$ (blue, full line), $\We=5$  (green, dashed line) and $\We=10$ (red, dotted line). Snapshots of the drop shape at t=0,2,4 ... for $\We=5$ are shown to the right of the graphs.  (c) Poincar\'e section for the non-axisymmetric case in the $a=1$ plane showing $\dot{a}$ against the potential energy $U$. The colours denote different starting points in the phase space.}
\end{figure*}

 \begin{figure*}
    \centering
  \includegraphics[width=0.55\linewidth]{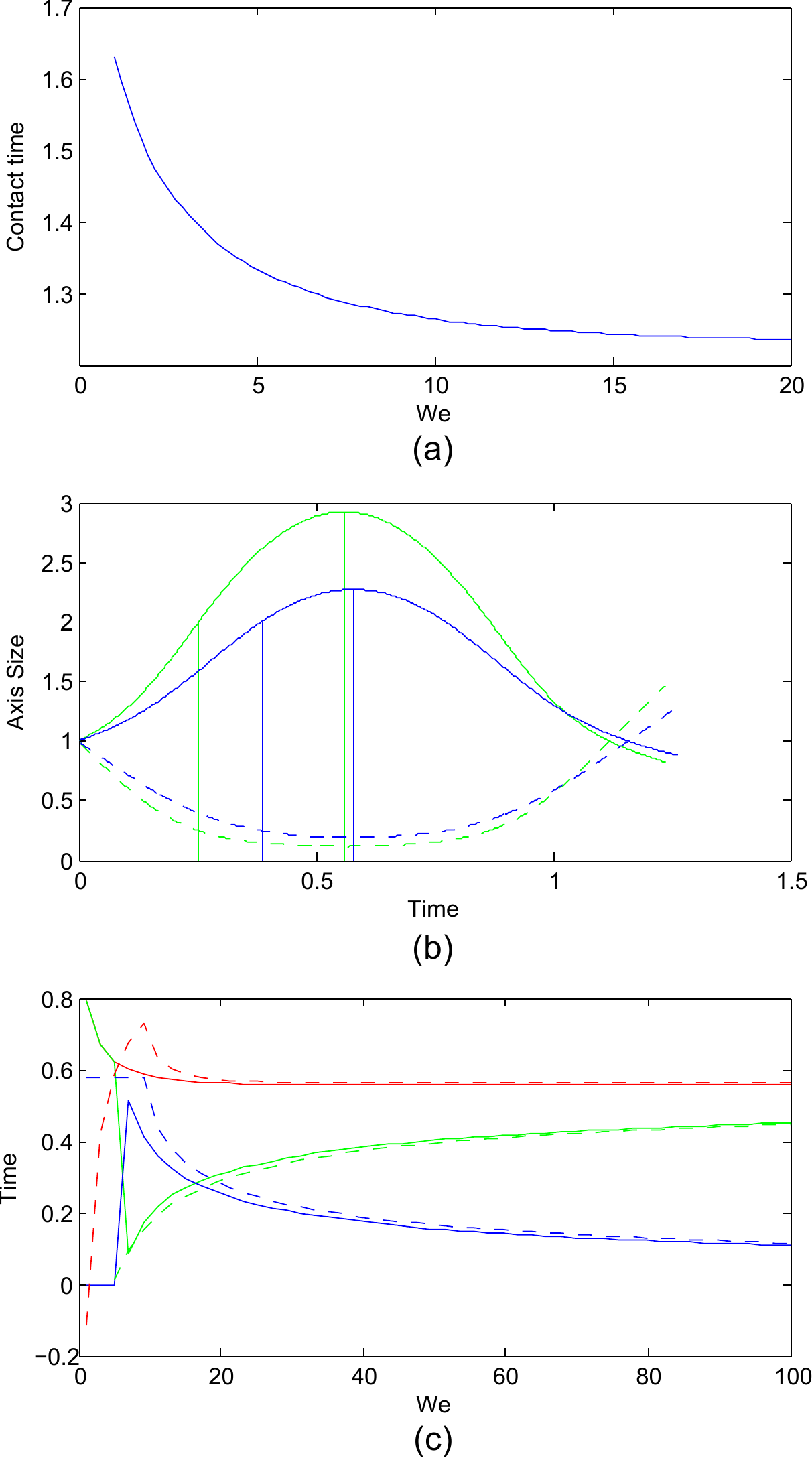}
     %\centering
 % \includegraphics[width=0.6\linewidth]{fig3b}
  \caption{\label{fig:fig3}(a) Variation of the contact time with Weber number for axisymmetric impacts. (b) Axis lengths, $c$: dotted, $a=b$ full lines, against time for $\We=10$ (blue) and $\We=20$ (green). The vertical lines show $t_1$ and $t_2$ for each $\We$. (c) A comparison of $t_1$ and $t_2$ for the numerical (solid lines) and analytical (dashed lines) results against Weber number with $t_1$ (blue), $t_2$ (green), $t_1+t_2$ (red).}
 \end{figure*}
 
\section{Results}
\label{sec3}

\subsection{Free drop oscillations}

%\begin{figure}[h]
%\centering
%\includegraphics[width=\linewidth]{fig1}
 %\centering
%\includegraphics[width=\linewidth]{fig2}
 %\centering
%\includegraphics[width=0.5\linewidth]{poincare6}
 %\caption{\label{fig:freeosc}(a) Time evolution of the vertical axis $c$ of a free drop oscillating axisymmetrically about $c$ at $\We=1 (blue)$, $\We=5$  (green) and $\We=10$ (red). Snapshots of the drop shape at t=0,1,2 ... for $\We=5$ are shown beneath the graph. (b) Time evolution of the vertical axis $c$ of a free drop oscillating asymmetrically ($\dot{a}\neq\dot{b}$) at $\We=1$ (blue), $\We=5$  (green) and $\We=10$ (red). Snapshots of the drop shape at t=0,5,10 ... for $\We=5$ are shown beneath the graph. (c) Poincar\'e section for the asymmetric case in the $a=1$ plane showing $\dot{a}$ against the potential energy $U$. The colours denote different starting points in the phase space. }
%\end{figure}

%\jucomment{exact solution?}

We first consider the oscillations of a free drop. The drop can be initialised in an axisymmetric mode by choosing $\gamma=1$. It then oscillates between an oblate and a prolate spheroid, as shown in Fig.~\ref{fig:freeosc}a.  For small amplitudes the model captures the Rayleigh result for the period of oscillation, as expected.  However, as the Weber number is increased, the drop no longer oscillates harmonically: in particular it oscillates far more quickly out of the oblate ellipsoid shape than it does from the prolate ellipsoid. Moreover, the period increases by approximately an order of magnitude as $\We$ increases from $1$ to $10$. 
(It should be noted that the extreme prolate shapes here are unphysical as drop breakup would occur.)

To initialise the drop in the non-axisymmetric mode we choose $\gamma =-1$. In line with the Rayleigh predictions for small amplitudes this mode has the same oscillation time as the axisymmetric mode. At higher amplitudes a pressure-mediated coupling between modes becomes important and the drop quickly starts to oscillates in a mixed mode shown in Fig.~\ref{fig:freeosc}b. Fig.~\ref{fig:freeosc}c shows a Poincare section of this mixed mode case. It can be seen that the motion is chaotic in nature.

% \begin{figure}
 %\includegraphics[width=\textwidth]{fig2}
 %\caption{\label{fig:asymsnap} (a) Evolution of $c$ of a free non-axisymmetric drop undergoing oscillations at We=1 (blue), We=5  (green) and We=10 (red). (b) Snapshots of drop shape evolution at time t=0,2,4,6,8,10. For We=5.}
 %\end{figure}

\subsection{Axisymmetric bouncing}

The case of axisymmetric bouncing, when the drop retains its circular shape during spreading and retraction, can be fully solved analytically.
In this case $a=b$, $c=1/a^2$ and the potential energy
\begin{equation}
U=U_1(a)=3^{1-1/\alpha} a^2 \lb 1 + 2 a^{-3\alpha}  \rb^{1/\alpha},
\label{e:U1}
\end{equation}
while the kinetic energy
\begin{equation}
T=m(a) \dot{a}^2, \quad m(a) = I \lb 1 + \frac{12}{a^6} \rb.
\label{e:T1}
\end{equation}
From the energy conservation law (\ref{e:E}) 
\begin{equation}
\dot{a}^2 = \frac{E-U_1(a)}{m(a)},
\label{e:dota}
\end{equation}
hence the time to expand between two energetically allowed droplet radii $a_1$ and $a_2$ is
\begin{equation}
t(a_1,a_2)= \int_{a_1}^{a_2} \sqrt{\frac{m(a)}{E-U_1(a)}} da.
\label{e:t}
\end{equation}
$a_1$ and $a_2$ are bounded by: 
\[ 
{a^{turn}_l \le a_1 \le a_2 \le a^{turn}_r},
\]
where the left and right turning points are determined by the condition 
\begin{equation}
U_1(a^{turn}_{i})=E, \quad i=l,r.
\label{e:turn}
\end{equation}
Since the Lagrangian dynamics is time-reversible, the limits of integration should be reversed for $a_2<a_1$. 

In order to find the moment of drop lift-off  we notice that $c=1/a^{2}$ and hence
\begin{equation}
\dot{c}= \dot{a} \frac{d}{da} \frac{1}{a^2}=-2 \frac{\dot{a}}{a^3} . \nonumber 
\end{equation}
Therefore we conclude that at the lift-off moment characterised by $\ddot{c}=0$ and $\dot{c} > 0$, the radial drop velocity is negative, $\dot{a}<0$, and the ratio $(\dot{a}/a^3)$ reaches its minimum. Taking into account (\ref{e:dota}), the take-off condition reads
\begin{equation}
\frac{d}{da} \lsb \frac{E-U_1(a)}{a^6 m(a)}  \rsb =0 \quad \mbox{or} \quad
\frac{d}{da} \lsb \frac{E-U_1(a)}{a^6 + 12}  \rsb =0.
\label{e:liftoff}
\end{equation}
It follows immediately from (\ref{e:liftoff}) that
both droplet inertia {\it and} surface tension are essential for lift-off. Indeed, assuming that the surface tension effects are unimportant  immediately leads to a contradiction as (\ref{e:liftoff}) can not be satisfied for $U_1 \equiv 0, \, a>0$. 
The explicit expression for the drop lift-off size $a_{\text{lift}}$ (\ref{e:liftoff}) can be found for small deviations from $a_{\text{lift}}=1$ by linearising (\ref{e:liftoff}). It results in 
\begin{equation}
a_{\text{lift}} \approx 1-\frac{E-3}{13 \alpha} = 1 - 0.024 \, \We.
\label{e:alift}
\end{equation} 
Thus, the lift-off occurs very soon after the drop radius $a$ has returned to the original size $1$. It is clear, therefore, that a suitable approximation for the time that the drop is in contact with the surface $\tau_{contact}$ for moderate ${\We}$ numbers is given by the half-oscillation duration ${\tau_{contact} \approx \tau_{1/2} = 2 \, t(1,a^{turn}_r)}$.

Fig.~\ref{fig:fig3}a shows the dependence of the drop contact time  on the Weber number obtained in numerical simulations of the model (\ref{e:Ia}),(\ref{e:Ib}),(\ref{e:V0}) and (\ref{e:Ic2}),  with the lift-off condition (\ref{cond:lift}). The contact time decreases with increasing $\We$, rapidly converging to its limiting value $1.24$; for $\We > 10$ the contact time becomes virtually independent of the impact velocity. A qualitatively similar behaviour has been observed in experiments\cite{richard2002surface}.  However the results are not a quantitative match as the measured contact time equals $2.6 \tau_e$ whereas the plateau for the model occurs at $1.24 \tau_s$ where  $\tau_e$ and  $\tau_s$  are the values of $(\rho R^3 / \sigma)^{\frac{1}{2}}$ for the experiment and simulation respectively. This is not unexpected because the model neglects many factors present in experiment, most notably the rim which tends to form around drops during retraction.

This dependence of the contact time on $\We$ finds an easy explanation within our model. 
Indeed, for large enough $\We$, upon collision with the surface the drop height $c$ quickly decreases and, correspondingly, the magnitude of drop spread $a$ becomes large. Hence, according to (\ref{e:U1}) and (\ref{e:T1}), the Lagrangian of the system can be approximated as 
\begin{equation}
\mathcal{L} \approx  I \dot{a}^2  - 3^{1-1/\alpha} a^2,
\label{e:Llarge} 
\end{equation}
for $a>a_c \sim 2$ (we shall return to discussing the more precise value for $a_c$ at the end of this section). This can be recognised as the Lagrangian for an oscillator of frequency 
\begin{equation}
\omega_l=\sqrt{ 3^{1-1/\alpha}/I  } \approx 2.75,
\label{e:omega_l}
\end{equation}
which is independent of the initial drop velocity. The corresponding half-oscillation period $\tau_{1/2} \approx 1.14$. For large $\We$ the drop contact time is dominated by the duration of drop spreading in this regime. The discrepancy with the limiting contact time value of $1.24$ is primarily due to lift-off occurring slightly after half an oscillation. 

The tendency of contact times to increase for smaller $\We$ can be traced to the longer periods of small (axisymmetric) oscillations. For $\Delta a = a -1 \ll 1$,
\begin{equation}
\mathcal{L} \approx  13 I \dot{(\Delta a)}^2  - 3 \lb 1+ \alpha  (\Delta a)^2 \rb, 
\label{e:Lsmall}
\end{equation}
and the corresponding frequency of oscillations is
\[
\omega_s= \sqrt{\frac{3 \alpha}{13 I}} \approx 1.36,
\]
with the half-oscillation period $\approx 2.31$. 

Thus, for large enough $\We$, the spreading dynamics of a bouncing drop is comprised of two distinct stages: the first stage, defined by $a<a_c$, of duration $t_1$, is characterised by fast evolution of the vertical (axial) drop thickness $c$ and is followed by the second stage, defined by $a>a_c$, of duration $t_2$, characterised by spreading mainly in the horizontal (radial) direction, see Fig.~\ref{fig:fig3}b. The minimal drop thickness $c_{min}$ is attained during the second stage; it is straightforward to show from the energy conservation law that $c_{min} \sim \We^{-1}$. The total half-oscillation period ${\tau_{1/2}=2(t_1+t_2)}$.

The half-oscillation period is dominated by $t_2$ only when $c \ll a$ for the majority of the oscillation, which is a reasonable assumption only for very high Weber numbers. Therefore it is surprising that, according to Fig.~\ref{fig:fig3}a, the contact time is already approaching its asymptotic value for $\We \sim 5$. To understand this further, we calculate $t_1$ for the small oscillations approximation of the Lagrangian (\ref{e:Lsmall}):
\begin{eqnarray}
	t_1=t(1,a_c) \approx \frac{1}{\omega_s} \arcsin{\frac{a_c-1}{a_s}}, \quad a_s=\sqrt{\frac{\We}{6\alpha}}. 
%\nonumber\\ t_1 \approx \sqrt{\frac{26 I}{\We}}(a_c-1), \quad \We >>1.
\label{e:t1}
\end{eqnarray}
Analogously, $t_2$ calculated using the large oscillations approximation of the Lagrangian (\ref{e:Llarge}) is
\begin{eqnarray}
t_2=t(a_c,a^{turn}_r) \approx  \frac{1}{\omega_l} \lb \frac{\pi}{2} - \arcsin{\frac{a_c}{a_l}} \rb, \quad a_l=\sqrt{ 3^{1/\alpha} (\We+6)  }. \nonumber
\label{e:t2}
\end{eqnarray}
For large $\We$,
\[
t_1 \approx \sqrt{\frac{26 I}{\We}}(a_c-1), \quad t_2 \approx \frac{\pi}{2\omega_l} - \sqrt{\frac{2I}{\We}}a_c.
\]
The value of the cut-off length $a_c$ should be chosen such that the sum $t_1+t_2$ only weakly depends on it. For $a_c \approx 1.3$ the two terms dependent on the Weber number cancel, hence the sum is only weakly influenced by the initial kinetic energy and the plateau in contact time is reached quickly with increasing $\We$. Fig.~\ref{fig:fig3}c shows how the approximations for $t_1$ and $t_2$ compare to the times measured in the simulations. 

%Fig.~\ref{fig:fig3}b indicates the variation of the axis lengths $a$ and $b$ with time during the oscillation for two different values of the Weber number. 

We 
point out for future reference that the turning point gives a relatively large contribution to the bouncing time as the integrand in (\ref{e:t}) diverges; a $10\%$ neighbourhood of the turning point contributes about $30\%$ of $\tau_{contact}$.

The major effect of including the contact line drag $F_a=F_b=F$ in the model is in breaking the time-reversal symmetry of axisymmetric drop spreading and retraction: the damping tends to decrease the spreading time and increase the retraction time. The interplay of these two effects decreases the contact time for $F<~1$ and increases it for higher values of $F$.

 \subsection{Non-axisymmetric bouncing} 

Several authors have recently shown that the drop-substrate contact time is reduced if the bouncing is not axisymmetric\cite{bird2013reducing, song2015selectively, liu2015symmetry, gauthier2015water}. Non-axisymmetric bouncing may result from an asymmetry in the initial conditions, such as different initial momenta along the $a$ and $b$ directions or a non-axisymmetric drop shape at the collision, or from anisotropy of the physical process of interaction of the drop with the substrate, such as anisotropic surface drag. In section~\ref{A} we give analytical arguments to show that, given an initial anisotropy, the drop dynamics may lead to development of strongly non-axisymmetric shapes. We discuss the roles of surface tension, pressure and inertia forces and link the shortening of contact times to the non-axisymmetric bouncing. In section~\ref{B} we use numerical solutions of the equations of motion to confirm and extend our conclusions.

\subsubsection{Analytical arguments: }

\label{A}

According to the governing equations  (\ref{e:Ia}), (\ref{e:Ib}), (\ref{e:V0}) and (\ref{e:Ic2}), the drop shape dynamics is determined by interplay of three forces: the drop inertia, surface tension and pressure. Our goal is to find which of these factors lead to the development of non-axisymmetric drop shapes. To this end, we subtract Eq. (\ref{e:Ib}) from Eq. (\ref{e:Ia}) and obtain
\begin{eqnarray}
&&I \frac{d^2}{dt^2}(a-b)+ R(a,b) (a-b)=0,\label{e:asym}\\
&&R(a,b)= \frac{\tilde{p}}{ab} + \frac{\partial_a U - \partial_b U}{a-b}. \label{e:R}
\end{eqnarray}
The difference $(a-b)$ measures the drop shape asymmetry.
If $R(a,b)>0$, both eigenvalues of the linearised Eq. (\ref{e:asym}) are imaginary and hence the local dynamics of $(a-b)$ is oscillatory. But if $R(a,b)<0$, one of the eigenvalues becomes real positive and the dynamics of $(a-b)$ is linearly unstable. Then, any asymmetry of the drop shape will grow exponentially. Having made this observation, we now turn to discussing the dependence of $R(a,b)$ on the physical parameters of the problem.

\begin{figure}[h]
\includegraphics[scale=0.5]{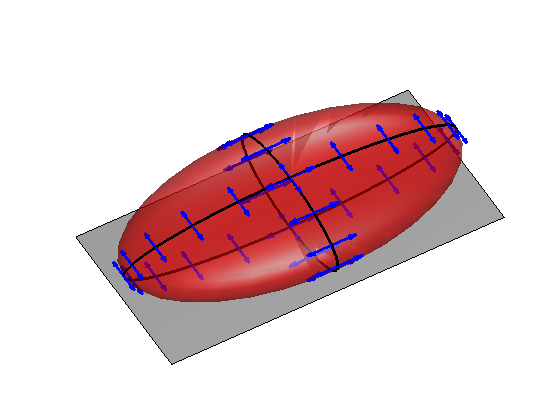}
\caption{Ellipsoidal drop. The tangential surface tension forces along the contour $a=\mbox{const}$ of length $l_a$ and the contour $b=\mbox{const}$ of length $l_b$. For $a>b$, $l_a<l_b$ and $F_a<F_b$, i.e. the longer horizontal axis of the drop experiences a lesser contractile force.}
\label{f:Ellipsoid}
\end{figure}

The first term on the right hand side of Eq.~(\ref{e:R}) describes the effect of pressure and therefore must be positive on physical grounds. Hence, it cannot lead to growth of the drop shape asymmetry. The second term on the right hand side of (\ref{e:R}) describes the effect of surface tension. It can be easily shown that it is negative and, therefore, will lead to drop shape asymmetry growth. Indeed, tangential surface tension forces acting along a closed contour are proportional to its length. Therefore, as illustrated in Fig.~\ref{f:Ellipsoid}, the total surface tension force $F_a$ ($F_b$) acting along the direction $a$ ($b$) is proportional to the length $l_a$ ($l_b$) of the contour lying in the plane $a=\mbox{const}$ ($b=\mbox{const}$). If $a>b$, $l_a<l_b$ and $F_a<F_b$, i.e.~the longer horizontal axis of the drop will experience a lesser contractile force. 

Eq.~(\ref{e:R}) shows that the character of the dynamics of the drop asymmetry is determined by the competition of the pressure and surface tension forces. In particular it is independent of the inertia forces which only affect the rate of the dynamics. 
%However before turning to  dynamics of anisotropic bouncing we would like to discuss what drop shapes are allowed by the energy conservation law.

Similarly to the axisymmetric case, for $\We > 5$, following a collision with the surface, the drop dynamics is usefully decomposed into two stages: the first stage is characterised by a quick flattening of the drop and it is followed by the second stage characterised by a slow evolution of $c(t) \ll 1$. It turns out that the drop spreading dynamics during the second stage lends itself to a considerably simplified description. In order to demonstrate this we use the relation $c=(ab)^{-1}$ and re-write the potential energy as
\begin{equation}
U=U_2(a,b)=3^{1-1/\alpha} ( (ab)^\alpha + a^{-\alpha} + b^{-\alpha} )^{1/\alpha}.
\label{e:U2}
\end{equation}
%Isolines of $U_2$ are superimposed on the contours of constant drop thickness $c=(ab)^{-1}=\mbox{const}$ (black dotted lines). 
%----
%\begin{figure}[h]
%\includegraphics[scale=0.55]{potentialU.png}
%\includegraphics[scale=0.45]{Qdrop_boundaries.png}
%\caption{(a) Potential energy $U=U_2(a,b)$. (b) Isolines of $U_2$ (blue lines) superimposed on the contours of constant drop thickness $c=(ab)^{-1}=\mbox{const}$ (black dotted lines).}
%\label{f:U2}
%\end{figure}
%----
%As $\We$ increases, isolines of potential energy quickly become coincident with the contour lines of drop thickness. This means that motions preserving drop thickness are energetically neutral. 
%This is one of the reasons for simplification of the drop dynamics in the surface-tension driven stage, which we turn to next. 

%The energy conservation law restricts trajectories to $U_2(a,b) \le E=\We/2+3$. It is easy to see that the minimal allowed drop thickness $c_{min}^*$ is reached in axisymmetric bouncing. For large enough $\We$ it scales as 
%\begin{equation}
%c_{min}^* \sim \We^{-1}.
%\end{equation}
%For non-axisymmetric bouncing $c_{min} \ge c_{min}^*$. The potential energy form also restricts the maximal allowed drop extension 
%\begin{equation}
%a,b \le l_{max}^* \sim {\We}^2
%\label{e:bound1}
%\end{equation}
%and the maximal allowed drop elongation 
%\begin{equation}
%a/b \le f_{max}^* \sim \We^3.
%\label{e:bound2}
%\end{equation}
%The bounds (\ref{e:bound1}) and (\ref{e:bound2}) are unlikely to be reached by an actual trajectory but they indicate the character of surface tension forces.
%----
\begin{figure}[h]
\includegraphics[scale=0.5]{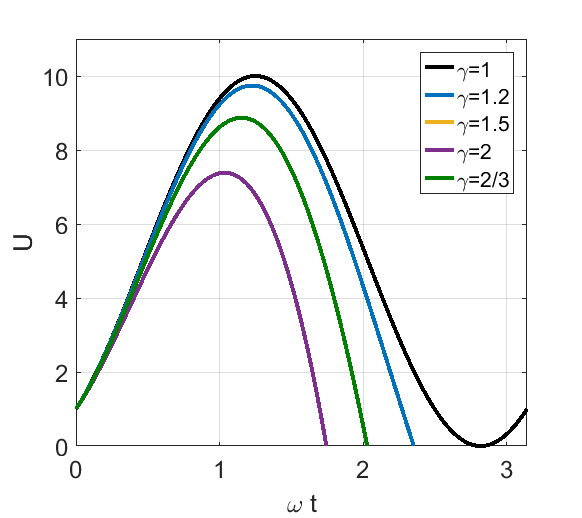}
\caption{Evolution of the potential energy $U$ for varying impact anisotropy $\gamma$. The drop retracts sooner for increasing $\gamma$. The maximal potential energy $U_{max} = E$ is reached only for $\gamma=1$ and decreases with increasing impact anisotropy. The non-axisymmetrically bouncing drop never comes to a full standstill. (Note that $U(t)$ is symmetric with respect to $\gamma \to \gamma^{-1}$, hence the curves for $\gamma=2/3$ and $\gamma=1.5$ coincide.)}
\label{f:Uanisotr}
\end{figure}
%-----
For $c \ll a, \, b$, to leading order, 
%During the surface tension-dominated stage of drop spreading and retraction, 
% $c \ll a, \, b$ and the dynamical equations for $a$ and $b$ decouple from the dynamical equation for $c$. Indeed, to the leading order 
\begin{eqnarray}
T \approx \frac{I}{2} (\dot{a}^2+\dot{b}^2), \quad U \approx 3^{1-1/\alpha} ab
\label{e:TUapprox_asym}
\end{eqnarray}
and the dynamical equations for $a$ and $b$ decouple from the dynamical equation for $c$:
\begin{eqnarray}
\ddot{a} = - \omega_l^2 b, \quad 
\ddot{b} = - \omega_l^2 a,  \label{e:ab}
\end{eqnarray}
with $\omega_l$ defined by Eq.~(\ref{e:omega_l}). In this approximation pressure plays no role in the dynamics of $a$ and $b$; hence, we should expect that development of drop shape anisotropy is most pronounced during this stage of drop spreading and retraction. (By contrast, to construct the proper approximation for the dynamics of $c$ one needs to go beyond the leading order approximation in (\ref{e:TUapprox_asym}). Then the pressure term emerges as the principal factor determining the dynamics of $c$.) 

The dynamics described by Eqs.~(\ref{e:ab}) is no longer oscillatory; indeed they can be immediately solved to produce 
\begin{eqnarray*}
a(t)&=& C_1 \sinh{\omega_l t} + C_2 \cosh{\omega_l t} + C_3 \sin{\omega_l t} + C_4 \cosh{\omega_l t}, \\ 
\nonumber
b(t)&=& -C_1 \sinh{\omega_l t} - C_2 \cosh{\omega_l t} + C_3 \sin{\omega_l t} + C_4 \cosh{\omega_l t},
\end{eqnarray*}
where the coefficients $C_k, \, k=1,...,4$ are related to the initial conditions at the beginning of the second stage as
\begin{eqnarray*}
C_1=\frac{\dot{a}_0-\dot{b}_0}{2 \omega_l}, \quad C_2=\frac{a_0-b_0}{2}, \quad
C_3=\frac{\dot{a}_0+\dot{b}_0}{2 \omega_l}, \quad C_4=\frac{a_0+b_0}{2}.
\end{eqnarray*}
For collisions leading to an anisotropic distribution of momentum in an initially axisymmetric drop: 
\begin{eqnarray*}
C_1=\sqrt{\We} \frac{\gamma-1}{\sqrt{1+\gamma^2}}, \quad
C_2=0, \quad C_3=\sqrt{\We} \frac{\gamma+1}{\sqrt{1+\gamma^2}}, \quad
C_4=a_c.
\end{eqnarray*}
Analogously, for axisymmetric impacts of a drop having an anisotropic initial shape:
\begin{eqnarray*}
C_1=0, \quad C_2=\frac{f-1}{2f}a_0, \quad
C_3=\sqrt{2 \, \We}, \quad
C_4=\frac{f+1}{2f}a_0
\end{eqnarray*}
where $f=b_0/a_0$.

Figure \ref{f:Uanisotr} shows the evolution of the potential energy $U(t)$ for varying impact anisotropy $\gamma$. Clearly, the drop retracts sooner for increasing impact anisotropy. The potential energy $U$ reaches the total energy $E$ only for axisymmetric impacts and a non-axisymmetrically bouncing drop never comes to a full standstill. Hence, the singularity which strongly contributes to the contact time in the symmetric case is circumvented and the total contact time decreases. A similar conclusion holds for non-axisymmetric bouncing driven by an initial drop shape anisotropy. 

The current treatment has a number of limitations: most notably, it can not predict drop lift-off since this process involves interplay of both pressure and surface tension forces. Also, the assumptions $c \ll a, b, \; c \ll 1$ may be violated for quickly growing drop asymmetry leading to strong contraction of one of the axes. In order to overcome these limitations we now turn to numerical solutions of the Lagrangian model. 

\subsubsection{Numerical integration of the equations of motion: }
\label{B}
%\begin{eqnarray}
%\pda U \approx \frac{U}{a} \approx 3^{1-1/\alpha} b, \\
%\pdb U \approx \frac{U}{b} \approx 3^{1-1/\alpha} a, \\
%\end{eqnarray}
 
\begin{figure*}
  \includegraphics[width=1.0\linewidth]{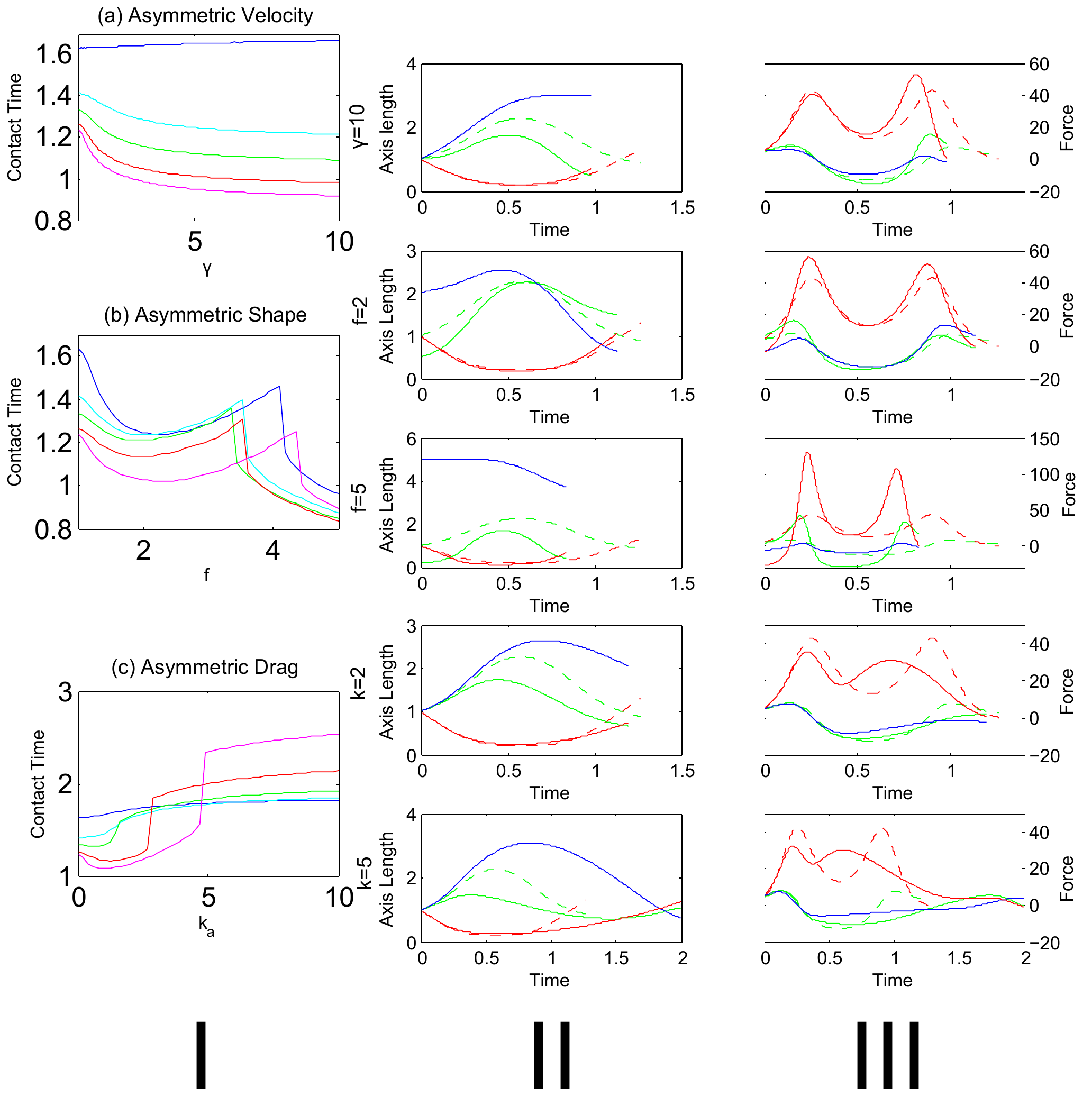}
  \caption{\label{f:aniso} Effect of anisotropy on the bouncing. \textbf{I} contact time as a function of asymmetry for $\We=1$ (blue), $\We=3$ (cyan), $\We=5$ (green), $\We=10$ (red) and $\We=20$ (magenta).  \textbf{II} variation in axis length a(green), b(blue) and c(red) with time for $\We=10$, solid lines are the non-axisymmetric case with dotted lines the symmetric case for comparison. \textbf{III} variation in force on axis a(green), b(blue) and c(red) with time for $\We=10$, solid lines are the non-axisymmetric case with dotted lines the symmetric case for comparison.}
\end{figure*}

 Our numerical results for different ways of breaking the axial symmetry are presented together in Fig.~\ref{f:aniso} in order to allow their comparison. For each case we show the variation of the contact time with anisotropy for different Weber numbers in panel I. We then choose $\We=10$ as an example and, for each case, show how the lengths of the axes and the forces acting on them vary with time, in panels II and III respectively. 
\paragraph{Anisotropic momentum. }
We first consider the effect of an anisotropic momentum distribution upon the collision of an axisymmetric drop with a flat surface. Therefore,  we impose an initial lateral asymmetry in momentum by taking $\gamma$ greater than unity, corresponding to $\dot{b}_0>\dot{a}_0$. Fig.~\ref{f:aniso}a(I) shows that, except for very small $\We \sim 1$, the contact time substantially decreases with increasing anisotropy, and that this effect is more pronounced at larger $\We$. 

Fig.~\ref{f:aniso}a(II) shows the dynamics of the drop. The initial anisotropy in momentum means that $b$ expands faster than $a$. 
This leads to a contractile surface tension force on $a$ that is larger than the one on $b$ (see Fig.~\ref{f:aniso}(III)), in accordance with the argument given in section A. Hence, $a$ reaches a maximum, and then starts to retract while $b$ is still growing. The drop shape anisotropy  at this stage is growing approximately exponentially with the rate $\omega_l$. Once $a$ starts to retract the incompressibility condition leads to a positive feedback which tends to slow down the oscillation of $b$. This feedback becomes more pronounced as $a$ grows shorter and drives the development of the drop anisotropy further. As $a$ becomes shorter and $b$, driven by surface tension, slows down and reaches a maximum, pressure increases and causes the total force on $c$ to increase. Hence, the center of mass of the drop attains a positive vertical velocity $\dot{c}$. The drop starts to expand in the vertical direction and this expansion eventually drives the pressure force down. Hence, the combined force on $c$ decreases and, finally, reaches zero. At this point the drop lifts off the surface. 

Note that, similar to the experiments\cite{liu2015symmetry} and in contrast with the axisymmetric drop dynamics, most of the change in contact time occurs during the retraction rather than the expansion stage.

For $\We=1$ there is a small increase in contact time. This occurs because the energy is insufficient for 
lift-off upon the initial retraction of the $a$-axis. The drop bounces at a later time as the $b$-axis retracts.

%Fig.~\ref{f:aniso}a(i) indicates that for $\We=1$ there is a small increase in contact time. This occurs because there is not sufficient energy for 
%lift-off upon the initial retraction of the the $a$-axis. The drop bounces at a later time as the $b$-axis retracts.  \\
%Looking at the $y$-axis the spreading and retraction are not mirror images of each other. You instead find that the spreading time changes slowly with asymmetry  as $t1$ decreases with increasing $\gamma$ as the initial velocity along this axis decreases, whereas $t_2$ increases with $\gamma$ as the surface tension force upon the $y$-axis Increases. The retraction time by contrast decreases substantially as it is driven by the surface tension force.  

\paragraph{Anisotropic shape. }
The impacts of drops with non-axisymmetric initial shapes also lead to changes in contact time. The mechanisms responsible for the drop bouncing dynamics are similar to those for initial momentum anisotropy. However they give rise to a more complicated dependence of the contact times on the initial shape.

%The drops were initialised with a fixed initial volume $V_0=4\pi/3$ and $c_0=1$, and the asymmetry was controlled by $f=b_0/a_0$.

Fig.~\ref{f:aniso}b shows a non-monotonic variation of the contact time on the parameter $f=b_0/a_0$, controlling the initial drop shape anisotropy. The physical difference between the bouncing for $f<\sim4$, to the left of the cusp, and for higher values of $f$ is that in the former case it is the retraction of the initially longer axis $b$ that drives the drop lift-off, while in the latter case it is the retraction of the initially shorter axis $a$.

For $f>\sim4$ the surface tension force acting on $a$ is initially large and therefore $a$ oscillates more quickly than $b$. Once it starts retracting the incompressibility condition leads to a coupling which further slows the oscillation of the $b$ axis and in turn promotes a faster $a$-retraction leading to quicker bouncing. This mechanism is fully analogous to that for the case of initial momentum anisotropy. 

%For $f<\sim4$, to the left of the cusp, the $b$ (initially larger) axis drives the lift-off, whereas for higher $f$ the $a$-axis retraction is responsible for the bouncing. For the larger anisotropies the bouncing mechanism is similar to that for the case where the velocities are initially anisotropic. The surface tension force on $a$ is larger and therefore it oscillates more quickly. Once it is retracting the incompressibility condition leads to a coupling which further slows the oscillation of the $b$ axis and in turn promotes anisotropy and fast $a$-retraction leading to quicker bouncing.

For smaller anisotropy, $f<\sim4$, the two directions are more balanced. $a$ still tends to oscillate more quickly but also to extend further before retracting. Hence the $b$ axis has ample time to contract first, and it drives the bouncing. In this case the shape anisotropy upon lift-off, and hence the reduction in contact time, are relatively small.\\

\paragraph{Anisotropic surface drag. }
Finally we consider the effect of anisotropic contact line drag on the bouncing of an initially axisymmetric drop. We assume that drag acts only on the moving $a$-axis, i.e. ${F_a=-k b \dot{a}, \; F_b=0}$. Fig~\ref{f:aniso}c(I) shows that here too there is a 
non-monotonic variation of contact time with $\We$. For low $k$ the bouncing mechanism is similar to that already described for anisotropic velocities, with the additional complication that for higher drag the slowing of the retraction due to the damping starts to have an effect.
For higher $k$ there is a different regime in which the contact time is greater than that for zero drag. This occurs when the damped $a$-axis retracts with insufficient energy to drive lift-off.

%\begin{comment}
%\begin{figure}
%  \includegraphics[width=0.5\linewidth]{symdrag}
%  \caption{\label{fig:fig5}(a) Contact time against initial shape asymmetry for various Weber numbers. (b) Variation in axis size for an symmetric and non-axisymmetric impact. (c) Pressure and surface tension force for both symmetric and asymmetric impacts.}
%\end{figure}
%\end{comment}

\section{Summary}

We have defined a simple Lagrangian model which is able to reproduce many of the features of the impact of drops on solid surfaces. The model extends the classic normal mode analysis of Rayleigh beyond the linear regime.  Our model qualitatively matches experiments on axisymmetric drop impact in that it shows a contact time that decreases to a plateau with increasing $\We$. The plateau occurs because the spreading and retraction is predominantly a simple harmonic motion driven by surface tension\cite{reyssat2010dynamical}. Quantitative difference between experiment and model are to be expected, because physical drops develop a rim upon bouncing, and because of viscous losses.

%The increase in contact time for low Weber numbers is also supported by experiments. One way in which it is not a realistic model of drop impact is that it�s shape is always constrained to be an ellipse, whereas in a real drop impact a rim film system is formed whereby the drop consists of a thin central film surrounded by a large rim containing most of the mass. It is worth noting that beyond the low weber number limit the driving forces in both these systems look similar however as both end up with a retraction force which is proportional to the area of a central ellipsoid shaped region. For the symmetric case this force is then merely proportional to the circle's radius leading to simple harmonic like force in both systems. This then explains why in both cases the dependence of the contact time upon the Weber number falls off once the impacts have enough energy to reach this state. 

We use the model to describe non-axisymmetric bouncing, due to an anisotropic initial velocity, initial shape or contact line drag. The usual effect of anisotropy is to cause a reduction in contact time. We show analytically that this occurs because once the drop has an elliptical footprint the surface tension force acting on the longer sides (or, equivalently, the direction perpendicular to the smaller initial velocity) is greater. Therefore the shorter axis retracts faster and, due to the incompressibility constraints, pumps fluid along the more extended droplet axis. This leads to a positive feedback, allowing the drop to jump in an elongated configuration, and more quickly.

This is the same as the mechanism described in Liu \textit{et al.}\cite{liu2015symmetry}, for drops bouncing on cylinders with radius larger than the drop radius, with the proviso that the physical drops develop a pronounced elevated rim during retraction which is not reproduced by the simple model considered here. For drops which bounce on smaller obstacles\cite{bird2013reducing} the reduction in contact time is due to drop break-up which is not included in our model.

\bibliography{mybib}{}
\bibliographystyle{rsc}
\end{document}